\documentclass[conference]{IEEEtran}
\usepackage{cite}
\usepackage{amsmath,amssymb,amsfonts}
\usepackage{algorithmic}
\usepackage{algorithm}
\usepackage{graphicx}
\usepackage{textcomp}
\usepackage{xcolor}
\def\BibTeX{{\rm B\kern-.05em{\sc i\kern-.025em b}\kern-.08em
    T\kern-.1667em\lower.7ex\hbox{E}\kern-.125emX}}
\begin{document}

\title{SB-RF: Schrödinger Bridge-based Rectified Flow for One-Step Robust Speech Enhancement
}

\author{Caixia~Lu,~Xueyang~Lv,~Penglong~Hu,~and~Jiaming~Xu
\thanks{All authors are with Xiaomi Corporation, Beijing, China. E-mail: lucaixia@xiaomi.com.}}

\maketitle

\begin{abstract}
Generative models have shown promising results for speech enhancement (SE), but they often rely on multi-step inference, limiting low-latency deployment. We propose SB-RF, a one-step generative framework that integrates Rectified Flow (RF) with Schrödinger Bridge (SB) theory. During training, SB-RF samples intermediate states from an SB time marginal and trains a conditional velocity field with the RF velocity-matching objective. At inference, SB-RF starts from the noisy observation and applies a single Euler update. Experiments show that SB-RF achieves competitive performance among generative methods on the VoiceBank-DEMAND benchmark. To further assess performance beyond this standard setting, we evaluate SB-RF on a simulated low signal-to-noise ratio test set using an expanded training dataset. Under these conditions, SB-RF achieves superior performance over the compared baselines, supporting its potential for real-world applications.
\end{abstract}


\section{Introduction}
Speech enhancement (SE) aims to recover clean speech from noisy observations, serving as a critical front-end for both human auditory perception (e.g., telecommunications, hearing aids) and downstream tasks (e.g., automatic speech recognition, speaker verification). 
Classical approaches such as spectral subtraction and Wiener filtering are computationally efficient but often struggle in low signal-to-noise ratio (SNR) scenarios. Data-driven approaches include discriminative models that learn a direct mapping from noisy to clean speech signals \cite{luo2018conv,lu2023mp,williamson2017time}, and generative models that learn the clean speech distribution and reconstruct the clean target conditioned on the noisy observation \cite{pascual2017segan,subramanian2019sergan,fang2021variational,leglaive2018variance}.

Although diffusion-based generative models have improved perceptual quality in SE, their slow inference due to the large number of function evaluations (NFE) remains a major limitation. Diffusion probabilistic models such as CDiffuSE \cite{lu2022conditional} formulate SE as an iterative reversal of a stochastic process, requiring 50--200 NFE. Continuous-time stochastic differential equation (SDE) approaches, such as SGMSE+ \cite{richter2023speech} and Brownian Bridge with Exponential Diffusion Coefficient (BBED) \cite{schneider2023reducing}, reduce NFE to 10--60.
Alternatively, SB-VE \cite{jukic2024schrodinger} directly applies Schrödinger Bridge theory to predict the clean signal, achieving high perceptual quality but still requiring 50 NFE. These results motivate the development of few-step and one-step generative SE methods. Flow Matching (FM) methods, including Rectified Flow (RF) \cite{liu2023flow}, learn a deterministic velocity field defined via an ordinary differential equation (ODE), achieving competitive results with NFE around 5 \cite{lee2025flowse}. 
Existing one-step generative SE approaches mainly follow two training strategies: distillation-based approaches \cite{voicerestore2025,flowingstraighter2024} train a student network to replicate the output of a multi-step teacher in a single forward pass, while direct one-step training methods such as LARF \cite{larf2024} and COSE \cite{compose2024} optimize the velocity field end-to-end for single-step inference. Recently, MeanFlowSE \cite{chen2025meanflowse} has been proposed as a one-step approach with strong reported performance. However, its performance relies heavily on large-scale pretrained encoders (e.g., WavLM-Large), resulting in significantly higher computational and memory costs that limit on-device deployment.

Standard RF constructs a straight-line interpolation path between paired clean and noisy speech. For each sample pair and timestep, this path provides only one training input to the velocity network. 
Training on such a limited set of inputs can make the learned velocity field sensitive to approximation errors and the mismatch between training and inference states during low-NFE inference, where only one or a few velocity evaluations are available and prediction errors are difficult to correct.
This limitation is particularly relevant to SE, where diverse speakers, phonetic content, SNR levels, and noise types require the velocity field to generalize beyond regions covered by the deterministic training path.

In this work, we propose Schrödinger Bridge-based Rectified Flow (SB-RF), a one-step generative framework that couples Schrödinger Bridge (SB) theory with RF. We first interpret standard RF as the deterministic (vanishing-variance) limit of a Brownian bridge, which motivates reintroducing stochasticity during training by sampling intermediate states $\mathbf{x}_t$ from the perturbation kernel of the Brownian bridge (BB) process. This perspective connects RF training with noise-perturbed learning used in denoising score matching and score-based generative modeling \cite{bishop1995tikhonov,vincent2011smdae,song2021scorebased}. 
We then replace this BB perturbation kernel with the analytic SB time marginal, which defines a stochastic interpolation between the clean-speech and noisy-observation endpoint marginals relative to a reference diffusion process \cite{leonard2012from,de2021diffusion,chen2018schrodinger,jukic2024schrodinger}. Finally, we retain the RF velocity-matching objective and perform one-step Euler inference. 

We evaluate the proposed method on the standard VoiceBank-DEMAND (VB-DMD) benchmark and a challenging low-SNR test set to assess robustness and generalization. Experiments demonstrate that SB-RF outperforms the compared baselines at NFE = 1, indicating the potential use of one-step generative SE in practical applications.
\section{Background}
The SE task is formulated in the Short-Time Fourier Transform (STFT) domain. Let $\mathbf{y}(f,k)$ denote the observed noisy speech spectrogram sampled from the distribution $\pi_y$ and $\mathbf{x}(f,k)$ denote the clean speech spectrogram sampled from the distribution $\pi_x$, where $f \in \{1, \dots, F\}$ and $k \in \{1, \dots, K\}$ are the frequency bin and time frame index, respectively. The observation signal in the single-channel setting is given by:
\begin{equation}
    \mathbf{y}\left(f,k\right) = \mathbf{x}\left(f,k\right) + \mathbf{n}\left(f,k\right),
    \label{eq:stft_model}
\end{equation}
where $\mathbf{n}(f,k)$ denotes the additive noise.
\subsection{Rectified Flow for Speech Enhancement}
\label{sec:Rectified Flow}
RF is a transport-based generative framework that learns an ODE-driven deterministic velocity field to transport samples between two distributions \cite{larf2024}. Let $\mathbf{x}_t$ denote the intermediate state at continuous time $t \in [0, 1]$. Its evolution is governed by:
\begin{equation}
    \frac{\mathrm{d}\mathbf{x}_t}{\mathrm{d}t} = \mathbf{v}_\theta(\mathbf{x}_t, t, \mathbf{y}),\quad\mathbf{x}_0=\mathbf{x}, \mathbf{x}_1=\mathbf{y},
    \label{eq:ode_dynamics}
\end{equation}
where $\mathbf{v}_\theta$ is a neural velocity field conditioned on $\mathbf{y}$, $t$ and $\mathbf{x}_t$. RF defines a deterministic trajectory via linear interpolation between $\mathbf{x} \sim \pi_x$ and $\mathbf{y} \sim \pi_y$:
\begin{equation}
    \mathbf{x}_t = t\mathbf{y} + (1-t)\mathbf{x}.
    \label{eq:interpolation}
\end{equation}
The velocity field $\mathbf{v}_\theta$ is trained to minimize the flow matching objective:
\begin{equation}
    \mathcal{L}_{v}(\theta) = \int_0^1 \mathbb{E}_{(\mathbf{x},\mathbf{y},\mathbf{x}_t)} \left[ \lVert (\mathbf{y} - \mathbf{x}) - \mathbf{v}_\theta(\mathbf{x}_t, t, \mathbf{y}) \rVert_2^2 \right] \mathrm{d}t.
    \label{eq:rf_loss}
\end{equation}
Although the straight-line assumption reduces NFE, training only on states from the deterministic path \eqref{eq:interpolation} exposes the velocity network to a limited set of training states.
\subsection{Schrödinger Bridge}
\label{sec:schrodinger_bridge}
While RF enforces a predefined linear interpolation, SB constructs a stochastic interpolation between two endpoint marginal distributions $\pi_x$ and $\pi_y$. Following SB formulations for generative modeling and speech enhancement \cite{de2021diffusion,jukic2024schrodinger,de2022likelihood}, we formulate SB as minimization of Kullback-Leibler (KL) divergence between a path measure $p$ and a reference path measure $p_{ref}$ with boundary conditions:
\begin{equation}
    \min_{p \in \mathcal{P}_{[0,T]}} D_{\mathrm{KL}}\left(p\,\lVert p_{ref}\right) \quad \text{s.t.} \quad p_0 = \pi_x, \ p_T = \pi_y,
    \label{eq:sb_kl_min}
\end{equation}
where $\mathcal{P}_{[0,T]}$ denotes the space of path measures on $[0,T]$. Equation \eqref{eq:sb_kl_min} enforces endpoint marginal constraints while regularizing the path measure toward the chosen reference process $p_{ref}$ \cite{leonard2012from,de2021diffusion}. The reference process $p_{ref}$ is typically a standard diffusion process governed by a drift $\mathbf{f}(\mathbf{x},t)$ and diffusion coefficient $g(t)$. The solution to \eqref{eq:sb_kl_min} satisfies a pair of forward-backward SDEs \cite{de2021diffusion,de2022likelihood,jukic2024schrodinger}:
\begin{subequations}
\begin{align}
\mathrm{d}\mathbf{x}_t &= \left[ \mathbf{f}(\mathbf{x}_t,t) + g^2(t) \nabla \log \Psi_t \right] \mathrm{d}t + g(t) \mathrm{d}\mathbf{w}_t, \mathbf{x}_0 \sim \pi_x, \\
\mathrm{d}\mathbf{x}_t &= \left[ \mathbf{f}(\mathbf{x}_t,t) - g^2(t) \nabla \log \bar{\Psi}_t \right] \mathrm{d}t + g(t) \mathrm{d}\bar{\mathbf{w}}_t, \mathbf{x}_T \sim \pi_y,
\end{align}
\end{subequations}
where $\mathbf{w}_t$ and $\bar{\mathbf{w}}_t$ are standard Wiener processes, $\Psi_t$ and $\bar{\Psi}_t$ are Schrödinger potentials whose log-gradients appear in the forward and reverse drift correction terms (i.e., $\nabla \log \Psi_t$ and $\nabla \log \bar{\Psi}_t$). In this formulation, the stochasticity of $\mathbf{x}_t$ is governed by both the endpoint marginal constraints ($p_0=\pi_x$, $p_T=\pi_y$) and the reference-diffusion regularization \cite{de2021diffusion,chen2018schrodinger,jukic2024schrodinger}.

While directly solving the SB problem is generally intractable and typically has no analytic solution, closed-form solutions exist for special cases. The SB time marginal can be written as $p_t(\cdot)=\Psi_t(\cdot)\bar{\Psi}_t(\cdot)$. Under Gaussian boundary conditions \cite{chen2023schrodingerbridges,chen2018schrodinger,jukic2024schrodinger}, the corresponding conditional sampling distribution for each paired clean and noisy sample $(\mathbf{x},\mathbf{y})$ is given by:
\begin{equation} 
q_t(\mathbf{x}_t  \mid \mathbf{x},\mathbf{y}) = \mathcal{N}_{\mathbb{C}} \left( \boldsymbol{\mu}_\mathbf{x}(t), \sigma_\mathbf{x}^2(t) \mathbf{I} \right),
\label{eq:SB-xt}
\end{equation}
where $\mathcal{N}_{\mathbb{C}}$ denotes a complex Gaussian distribution. The mean $\boldsymbol{\mu}_\mathbf{x}(t)$ and variance $\sigma_\mathbf{x}^2(t)$ in \eqref{eq:SB-xt} are defined as: 
\begin{equation}
\boldsymbol{\mu}_\mathbf{x}(t) = w_x(t) \mathbf{x} + w_y(t) \mathbf{y}, \quad
\sigma_\mathbf{x}^2(t) = \frac{\alpha_t^2 \bar{\sigma}_t^2 \sigma_t^2}{\sigma_T^2},
\label{eq:SB-xt-mu-sigma}
\end{equation}
where $w_x(t) = {\alpha_t \bar{\sigma}_t^2}/{\sigma_T^2}$ and $w_y(t) = {\bar{\alpha}_t \sigma_t^2}/{\sigma_T^2}$ are weighting factors derived from the reference diffusion schedule. Specifically, the schedule parameters are defined as $\alpha_t = e^{\int_0^t f(\tau)\mathrm{d}\tau}$, $\sigma_t^2 = \int_0^t g^2(\tau)/\alpha_\tau^2\,\mathrm{d}\tau$, $\bar{\alpha}_t = \alpha_t/\alpha_T$, and $\bar{\sigma}_t^2 = \sigma_T^2 - \sigma_t^2$.

For speech enhancement, this SB formulation is attractive because the clean-speech and noisy-speech distributions can be used as the two endpoint marginals. In the Gaussian case above, the closed-form conditional marginal provides a tractable way to sample stochastic intermediate states, with perturbations confined to interior times because the variance vanishes at both endpoints \cite{chen2018schrodinger,jukic2024schrodinger}.

\section{Proposed method}
This section describes the proposed SB-RF framework. We first revisit RF from a Brownian Bridge perspective to motivate stochastic intermediate-state sampling during training. Then we introduce SB-RF, which samples intermediate states from the SB time marginal during training and retains the RF inference rule for one-step speech enhancement.
\subsection{Brownian Bridge-based Rectified Flow}
\label{sec:bbed_rf_relation}
As noted in Section~\ref{sec:Rectified Flow}, standard RF provides intermediate states $\mathbf{x}_t$ using deterministic linear interpolation and trains the velocity field with the target displacement $\mathbf{y}-\mathbf{x}$ \cite{larf2024}. This path is efficient for low-NFE ODE inference, but it provides only one training state for each pair and timestep. 
Motivated by this limited set of training states, we introduce stochastic perturbations around the deterministic interpolation path by sampling intermediate states from a Gaussian distribution with time-dependent variance. The model therefore observes both the ideal interpolation state and nearby stochastic intermediate states, improving the robustness of the learned velocity field when only one or a few inference steps are used.
Such noise-perturbed training is consistent with denoising score matching and score-based generative modeling, where models are trained on perturbed data over different noise levels \cite{bishop1995tikhonov,vincent2011smdae,song2021scorebased}.

We first analyze a Brownian Bridge-based variant of RF. Consider the BB process used in generative speech enhancement (e.g., BBED \cite{schneider2023reducing}), where the intermediate state $\mathbf{x}_t$ at time $t$ follows:
\begin{equation} 
    \mathbf{x}_t = (1-t)\mathbf{x} + t\mathbf{y} + \sigma_t\mathbf{z},
    \label{eq:bb_kernel}
\end{equation}
where $\mathbf{z} \sim \mathcal{N}_{\mathbb{C}} \left( 0, \mathbf{I} \right)$, and $\sigma_t$ denotes a time-dependent perturbation scale. The corresponding variance $\sigma_t^2$ has a single peak inside the interval and vanishes at $t=0$ and $t=1$. We adopt the specific variance parameterization derived from the Ornstein-Uhlenbeck process as detailed in \cite{schneider2023reducing}.  

The dynamics of this process follow a standard SDE formulation where the drift term is $\mathbf{f}(\mathbf{x}_t,\mathbf{y},t)={(\mathbf{y}-\mathbf{x}_t)}/{(1-t)}$ and the diffusion term is $g(t) = \sqrt{c} k^t$. Here $k=\sigma_{\text{max}}/\sigma_{\text{min}}$ and $c= \sigma_{\text{min}}^2 \cdot 2\log\left(\frac{\sigma_{\text{max}}}{\sigma_{\text{min}}}\right)$ are hyperparameters adopted from \cite{schneider2023reducing}. 
Crucially, as the variance $\sigma_t^2 \to 0$, the stochastic term vanishes, and the intermediate state $\mathbf{x}_t$ converges to the deterministic mean path $(1-t)\mathbf{x} + t\mathbf{y}$. Consequently, the drift term simplifies to:
\begin{equation} 
    \lim_{\sigma_t \to 0} \frac{\mathbf{y} - \mathbf{x}_t}{1-t} = \mathbf{y} - \mathbf{x}.
    \label{eq:rf_limit}
\end{equation}
From \eqref{eq:rf_limit}, we recover the constant velocity target $\mathbf{v}(\mathbf{x}_t, t) = {\mathrm{d}\mathbf{x}_t}/{\mathrm{d}t} = \mathbf{y} - \mathbf{x}$ that is independent of $t$, which matches the learning objective of RF. This correspondence motivates viewing RF as the deterministic ODE limit of this BB parameterization when the bridge variance vanishes.

This zero-variance limit motivates a BB-RF solution that samples $\mathbf{x}_t$ from the perturbation kernel of the BB process in \eqref{eq:bb_kernel} during training. It introduces stochastic intermediate states while keeping the RF velocity-matching target $\mathbf{y}-\mathbf{x}$ and the one-step inference rule unchanged. 
\subsection{Schrödinger Bridge-based Rectified Flow}
\label{sec:schrodinger_bridge_rectified_flow}
Unlike standard RF, both BB-RF and SB-RF sample $\mathbf{x}_t$ stochastically during training. As shown in Section~\ref{sec:bbed_rf_relation}, BB-RF uses the perturbation kernel of the BB process in \eqref{eq:bb_kernel}, whose mean follows the linear path $(1{-}t)\mathbf{x}+t\mathbf{y}$ \cite{schneider2023reducing}. Although this construction provides a valid stochastic interpolation for each paired sample, the mean weights for the clean speech $\mathbf{x}$ and noisy observation $\mathbf{y}$ are fixed to $(1{-}t)$ and $t$, respectively. Since the relationship between noisy and clean speech can vary with speech content, SNR, and noise type, the choice of intermediate-state distribution can affect the learned velocity field. Therefore, a fixed linear mean path may be restrictive for modeling such intermediate states.

SB-RF samples $\mathbf{x}_t$ from the analytic SB time marginal in \eqref{eq:SB-xt} and \eqref{eq:SB-xt-mu-sigma} during training. As described in Section~\ref{sec:schrodinger_bridge}, this marginal is derived from the SB formulation that minimizes KL divergence relative to a reference process under endpoint marginal constraints \cite{leonard2012from,de2021diffusion,jukic2024schrodinger}. In the Gaussian case, this formulation jointly determines the mean weights $w_x(t),w_y(t)$ and the covariance $\sigma_{\mathbf{x}}^2(t)\mathbf{I}$ used for sampling intermediate states. Unlike BB-RF, the mean weights of the clean speech and noisy observation are not restricted to the fixed linear weights $(1{-}t)$ and $t$, and the perturbation scale is defined within the same marginal.

Given these sampled states, the network is trained with the RF velocity-matching objective in \eqref{eq:rf_loss}. The SB time marginal determines the intermediate states used as training inputs, while RF determines the supervised velocity target as $\mathbf{y}-\mathbf{x}$ and uses the Euler update for inference. Therefore, SB-RF does not require iterative SB sampling at inference time; it starts from the noisy observation and applies the same Euler update in \eqref{eq:inference_ode} \cite{liu2023flow,larf2024}. This preserves the one-step inference efficiency of RF while improving the robustness of the learned velocity field by incorporating the SB time marginal into RF training.

\subsection{Training and Inference}
Algorithm~\ref{alg:sbrf_train} and Algorithm~\ref{alg:sbrf_infer} summarize the training and inference procedures. Let $(\mathbf{x}, \mathbf{y}) \sim (\pi_{\text{clean}}, \pi_{\text{noisy}})$ denote pairs of clean and noisy STFT spectrograms sampled from their respective distributions. During training, the timestep $t$ is sampled uniformly from $[\epsilon, T]$, where $\epsilon>0$ is a small constant for numerical stability and $T$ represents the maximum diffusion time. The RF velocity target remains the full endpoint displacement $(\mathbf{y}-\mathbf{x})$, regardless of the sampled timestep $t$.

The intermediate state $\mathbf{x}_t$ is sampled from the analytic SB time marginal defined in \eqref{eq:SB-xt}. Furthermore, we adopt the variance-exploding noise schedule $\sigma_t^2={c\left(k^{2t} - 1\right)}/{2 \log(k)}$ with $\alpha_t=1$, following the prior SB-based speech enhancement work in \cite{jukic2024schrodinger}. 
We employ NCSN++ \cite{richter2023speech} as the backbone to parameterize the velocity field $\mathbf{v}_{\theta}(\mathbf{x}_t, t, \mathbf{y})$. The neural network is trained to minimize the following composite loss: 
\begin{equation} 
\mathcal{L}(\theta) = \mathcal{L}_{v}(\theta) + \lambda_1 \mathcal{L}_{mel}(\theta) + \lambda_2 \mathcal{L}_{pesq}(\theta).
    \label{eq:sbrf_loss}
\end{equation} 
Here, $\mathcal{L}_{v}(\theta)$ represents the velocity matching loss guided by RF in \eqref{eq:rf_loss}. 
To further improve reconstruction fidelity and perceptual quality, we incorporate the multi-resolution Mel-spectrogram loss $\mathcal{L}_{mel}(\theta)$ \cite{li2024compression} and the PESQ-based loss $\mathcal{L}_{pesq}(\theta)$, weighted by hyperparameters $\lambda_1$ and $\lambda_2$, respectively. 
Since the network is trained with the RF velocity-matching target $\mathbf{y}-\mathbf{x}$, the estimated clean speech $\hat{\mathbf{x}}$ used in $\mathcal{L}_{mel}(\theta)$ and $\mathcal{L}_{pesq}(\theta)$ is calculated by a single reverse Euler step from time $t$ to $\epsilon$: $\hat{\mathbf{x}} = \mathbf{x}_t - (t-\epsilon) \cdot \mathbf{v}_\theta(\mathbf{x}_t,t,\mathbf{y})$. This ensures that the auxiliary losses directly supervise one-step reconstruction quality at each training timestep $t$.

For inference, we use the Euler ODE solver to recover the clean speech. Starting from $\mathbf{x}_T = \mathbf{y}$ and $t=T$, the iterative update rule is given by: 
\begin{equation} 
\mathbf{x}_{t-\Delta t} = \mathbf{x}_t - \mathbf{v}_\theta(\mathbf{x}_t, t, \mathbf{y})\Delta t,
\label{eq:inference_ode}
\end{equation} 
where $\Delta t = (T-\epsilon)/N$ and $N$ denotes the number of inference steps. Finally, we reconstruct the enhanced speech via inverse STFT using the estimated complex spectrogram.
\begin{algorithm}[t]
  \caption{SB-RF training procedure}
  \label{alg:sbrf_train}
  \begin{algorithmic}[1]                           
  \REQUIRE Training pairs $(\mathbf{x},\mathbf{y})$
  \FOR{each batch}    
    \STATE Sample timestep $t\sim\mathcal{U}([\epsilon,T])$
    \STATE Sample $\mathbf{z}\sim\mathcal{N}_{\mathbb{C}}(0,\mathbf{I})$
    \STATE Compute SB mean and variance using \eqref{eq:SB-xt}--\eqref{eq:SB-xt-mu-sigma}
    \STATE Generate $\mathbf{x}_t=\boldsymbol{\mu}_{\mathbf{x}}(t)+\sigma_{\mathbf{x}}(t)\mathbf{z}$
    \STATE Estimate $\mathbf{v}_{\theta}(\mathbf{x}_t,t,\mathbf{y})$
    \STATE Compute composite loss in \eqref{eq:sbrf_loss}
    \STATE Backpropagate to update $\mathbf{v}_{\theta}$
  \ENDFOR
  \ENSURE Optimized velocity field $\mathbf{v}_{\theta}$
  \end{algorithmic}
  \end{algorithm}

  \begin{algorithm}[t]
  \caption{SB-RF inference procedure}
  \label{alg:sbrf_infer}
  \begin{algorithmic}[1]
  \REQUIRE Noisy spectrogram $\mathbf{y}$, optimized velocity field $\mathbf{v}_{\theta}$, number of inference steps $N$
  \STATE \textbf{Initialization:} $\mathbf{x}_{T}=\mathbf{y}$, $t=T$
  \STATE \textbf{Set step size:} $\Delta t=(T-\epsilon)/N$
  \FOR{each reverse-time step}
    \STATE Compute the velocity vector $\mathbf{v}_{\theta}(\mathbf{x}_t,t,\mathbf{y})$
    \STATE Update $\mathbf{x}_{t-\Delta t}=\mathbf{x}_t-\mathbf{v}_{\theta}(\mathbf{x}_t,t,\mathbf{y})\Delta t$
  \ENDFOR
  \ENSURE Enhanced spectrogram $\hat{\mathbf{x}}$
  \end{algorithmic}
  \end{algorithm}
\section{Experiments}
\subsection{Datasets and Evaluation Metrics}
\subsubsection{Datasets}
To evaluate both standard denoising performance and low-SNR generalization, we organize our experiments into two tracks. 
Track A follows the standard VB-DMD benchmark for comparison with prior work. 
Track B focuses on performance under challenging low-SNR conditions designed to simulate difficult real-world scenarios. 
This track introduces an expanded training set, as limited datasets may constrain the generalization of generative methods in complex environments. Furthermore, we generate a simulated low-SNR test set, as robust denoising under such conditions remains a critical challenge for real-world applications.
\begin{itemize}
\item \textbf{Standard setting (VB-DMD):}
The VB-DMD training set comprises 11,572 utterances, generated by mixing clean speech from the VCTK dataset \cite{yamagishi2019vctk-corpus} with DEMAND noise signals \cite{thiemann2013demand} and two artificial noises (babble and speech-shaped) at SNRs of {0, 5, 10, 15} dB. The test set contains 824 utterances with SNRs of {2.5, 7.5, 12.5, 17.5} dB.
\item \textbf{Low-SNR robustness setting:}
For training, we randomly sampled 1000 hours of clean speech: 500 hours from WenetSpeech4TTS (Premium subset) \cite{ma2024wenetspeech4tts} and 500 hours from the Deep Noise Suppression Challenge 4 (DNS-4) clean speech corpus \cite{dubey2022icassp}.
The noise set includes the DNS-4 noise corpus and MUSAN \cite{snyder2015musan}. Mixtures are generated on-the-fly with SNRs uniformly sampled from $-10$ to $15$ dB. 
For evaluation, we randomly selected 10,000 utterances of clean speech from AISHELL-1 \cite{bu2017aishell1} and LibriSpeech \cite{panayotov2015librispeech}, mixed with noise clips from WHAM! \cite{wichern2019wham} at SNRs uniformly sampled from $-10$ to $0$ dB. 
Crucially, we ensure no overlap in speakers, utterances, or noise clips between the training and test sets. 
\end{itemize}
\subsubsection{Evaluation Metrics}
The performance is evaluated in terms of perceptual evaluation of speech quality (PESQ) \cite{itu2001pesq} for perceived quality, extended short-term objective intelligibility (ESTOI) \cite{bronkhorst2002intelligibility} for intelligibility, scale-invariant measures SI-SDR, SI-SIR, and SI-SAR \cite{leroux2019sdr}, and DNSMOS P.808 for nonintrusive quality estimation \cite{paliwal2020dnsmos}. We also report NFE to assess efficiency.
\subsection{Implementation Details}
\textbf{Experimental Setup.}
All speech signals are resampled to 16 kHz, and then transformed into the complex STFT representation with a window size of 510 and a hop size of 128, following \cite{richter2023speech}. We obtain $F=256$ frequency bins and concatenate $K=256$ consecutive frames to form the model input $\mathbf{X}\in \mathbb{C}^{F \times K}$. An amplitude transformation $\tilde{\mathbf{c}} = \beta \left| \mathbf{c} \right|^\alpha e^{j \angle \mathbf{c}}$ is applied to compensate for the heavy-tailed distribution of all complex STFT coefficients $\mathbf{c}$ with $\alpha=0.5$ and $\beta=0.33$ \cite{gerkmann2010empirical,braun2021consolidated}. For the bridge process, the time boundaries are set to $\epsilon=0.03$ and $T=0.97$.

We adopt NCSN++ as the backbone with 65.6M parameters and 265 GMACs \cite{richter2023speech}. The weighting hyperparameters are set to $\lambda_1=33$ and $\lambda_2=3$. The multi-resolution Mel-spectrogram loss follows the configuration in \cite{li2024compression}. 
We use the Adam optimizer with an initial learning rate of $0.0001$ and an exponential decay of 0.999 per epoch. The batch size is set to 16 per GPU. All models are trained from scratch with random initialization. Training is conducted on 8 NVIDIA RTX A800 GPUs.

\textbf{Baseline Systems.}
The proposed SB-RF is compared against representative baselines.
MP-SENet \cite{lu2023mp} is a representative discriminative method. SGMSE+ \cite{richter2023speech} and BBED \cite{schneider2023reducing} are score-based generative methods. SB-VE \cite{jukic2024schrodinger} is a generative method that directly predicts the clean signal based on SB theory. FlowSE \cite{lee2025flowse}, LARF \cite{larf2024} and COSE \cite{compose2024} are velocity-matching generative methods, where LARF (modified RF) and COSE (average-velocity flow matching) are representative one-step generation benchmarks. 

\textbf{Evaluation Setup.}
For Track A, the proposed methods are trained and evaluated solely on VB-DMD, and all baseline results are taken from the original papers or obtained using author-released checkpoints when available. For Track B, we retrain selected baselines and the proposed method on the expanded training set. LARF and COSE are excluded because their official implementations were not publicly available at the time of our experiments.
All retrained baselines are implemented with officially released code and recommended configurations. Since existing baselines differ in backbone architectures, training objectives, loss designs, and sampling steps, we evaluate SB-RF against them at the system level, treating each method as a complete SE system. 

In addition to this system-level comparison, we include three controlled analyses. First, we vary the number of inference steps for SB-RF to examine whether multi-step Euler inference improves performance. Second, BB-RF and SB-RF share the same backbone, velocity target, auxiliary losses, and inference rule, differing only in the intermediate-state sampling strategy. Third, the auxiliary-loss ablation keeps the same SB-RF sampling and inference pipeline while training with different loss configurations.
\subsection{Experimental Results}
We first report the results on Tracks A and B, followed by analyses of inference steps, SB-based sampling, spectrogram visualization, and auxiliary losses.

\noindent\textbf{Track A: evaluation on standard VoiceBank-DEMAND.}
Table~\ref{tab:vb_dmd_comparison} compares the performance of different methods on the VB-DMD Track A test set. 
With one-step generation, the proposed SB-RF achieves the highest PESQ (3.39) and SI-SDR (19.5 dB) among the compared generative SE approaches, and matches the best ESTOI (0.88).
It also outperforms multi-step generative baselines such as SGMSE+, BBED and SB-VE, which require 15 to 60 NFE.
Compared to FlowSE (conditional flow matching), which typically requires NFE = 5, SB-RF yields higher PESQ and SI-SDR with one-step inference.
Furthermore, SB-RF outperforms one-step generative baselines such as LARF and COSE by a large margin (e.g., +0.37 PESQ over COSE, +0.42 PESQ over LARF). These results show that the proposed integration of sampling from the SB time marginal with the RF objective provides effective one-step enhancement among the compared generative methods.
\begin{table}[htbp]
  \caption{Speech enhancement results on Track A test set}
  \label{tab:vb_dmd_comparison}
  \centering   
  \begin{tabular}{|l|c|c|c|c|c|c|}
  \hline         
  \textbf{Method} & \textbf{NFE} & \textbf{PESQ} & \textbf{ESTOI} & \textbf{SI-SDR} & \textbf{SI-SIR} &           
  \textbf{SI-SAR} \\
  \hline
  noisy & -- & 1.97 & 0.79 & 8.4 & 8.4 & -- \\
  \hline
  SGMSE+ & 15 & 2.80 & 0.86 & 17.2 & 26.9 & 17.9 \\
  BBED & 30 & 3.09 & \textbf{0.88} & 18.8 & 30.1 & 19.4 \\
  SB-VE & 50 & 2.91 & \textbf{0.88} & 19.4 & -- & -- \\
  FlowSE & 5 & 3.12 & \textbf{0.88} & 19.0 & \textbf{32.2} & 19.4 \\
  COSE & 1 & 3.02 & 0.87 & 19.3 & 31.7 & 19.8 \\
  LARF & 1 & 2.97 & 0.87 & 19.2 & 26.4 & \textbf{20.7} \\
  \hline
  BB-RF & 1 & 3.28 & 0.87 & 18.9 & 28.7 & 19.9 \\
  SB-RF & 1 & \textbf{3.39} & \textbf{0.88} & \textbf{19.5} & 30.0 & 20.1 \\
  \hline
  \end{tabular}
  \end{table}

\noindent\textbf{Track B: evaluation on low signal-to-noise ratio conditions.}
Table~\ref{tab:simu_comparison2} presents generalization performance on the simulated low-SNR test set for Track B. 
In this setting, SB-RF demonstrates robust performance, achieving the highest PESQ (2.56) and ESTOI (0.70) among the compared systems.
Under unseen low-SNR conditions, SB-RF outperforms the representative discriminative baseline MP-SENet in perceptual quality and intelligibility metrics (+0.47 PESQ, +0.04 ESTOI, +0.12 DNSMOS) while maintaining a comparable SI-SDR (10.4 dB vs. 10.5 dB).
While BBED achieves the highest DNSMOS (3.49) among all baselines, SB-RF is marginally lower (3.41) but obtains the highest PESQ and ESTOI with NFE = 1. However, given that BBED requires NFE = 30, SB-RF offers a favorable trade-off between inference latency and overall enhancement quality. These results suggest that SB-RF remains effective under the evaluated challenging low-SNR range.
\begin{table}[htbp]                 
  \caption{Speech enhancement results on Track B test set}
  \label{tab:simu_comparison2}
  \centering
  \begin{tabular}{|l|c|c|c|c|c|}
  \hline
  \textbf{Method} & \textbf{NFE} & \textbf{PESQ} & \textbf{ESTOI} & \textbf{SI-SDR} & \textbf{DNSMOS} \\
  \hline
  noisy & -- & 1.12 & 0.36 & $-5.4$ & 2.42 \\
  \hline
  MP-SENet & 1 & 2.09 & 0.66 & 10.5 & 3.29 \\
  \hline
  BBED & 30 & 1.83 & 0.62 & 7.8 & \textbf{3.49} \\
  SB-VE & 50 & 2.07 & 0.66 & 9.1 & 3.42 \\
  \hline
  BB-RF & 1 & 2.43 & 0.66 & 9.1 & 3.36 \\
  SB-RF & 1 & \textbf{2.56} & \textbf{0.70} & 10.4 & 3.41 \\
  SB-RF & 5 & 2.46 & \textbf{0.70} & \textbf{10.7} & 3.42 \\
  SB-RF & 10 & 2.44 & \textbf{0.70} & \textbf{10.7} & 3.43 \\
  \hline
  \end{tabular}
  \end{table}

\begin{figure*}[htbp]
\centering
\begin{minipage}{0.32\textwidth}
  \centering
  \includegraphics[width=\linewidth]{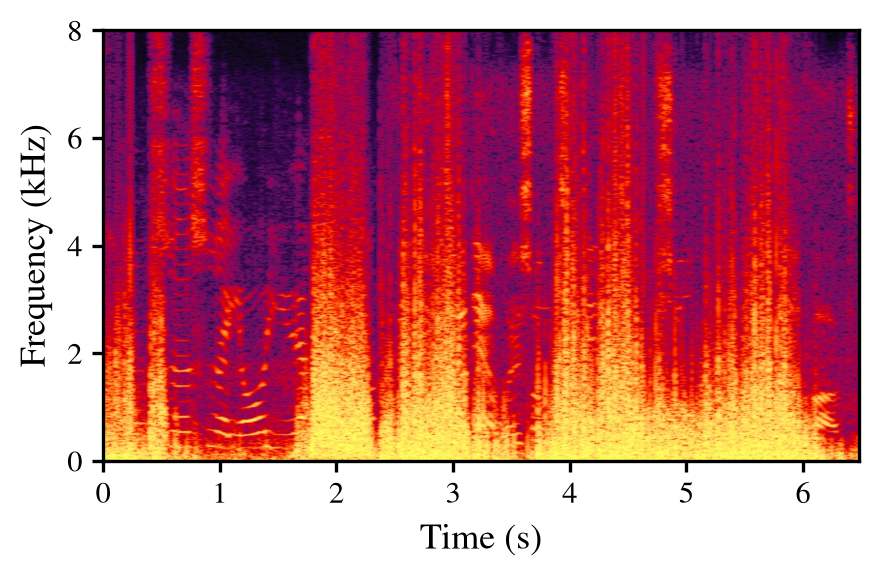}
  \centerline{\footnotesize (a) Noisy input}\medskip
\end{minipage}
\hfill
\begin{minipage}{0.32\textwidth}
  \centering
  \includegraphics[width=\linewidth]{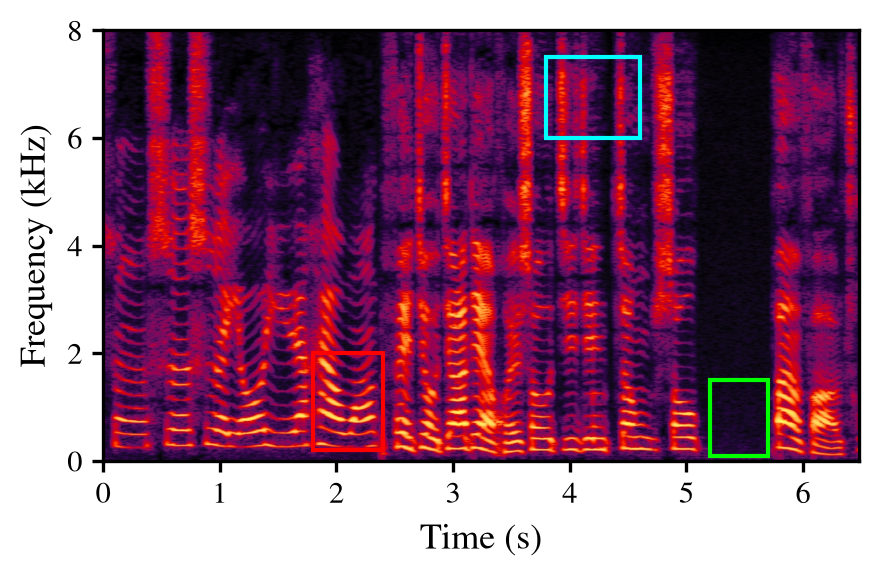}
  \centerline{\footnotesize (b) Clean reference}\medskip
\end{minipage}
\hfill
\begin{minipage}{0.32\textwidth}
  \centering
  \includegraphics[width=\linewidth]{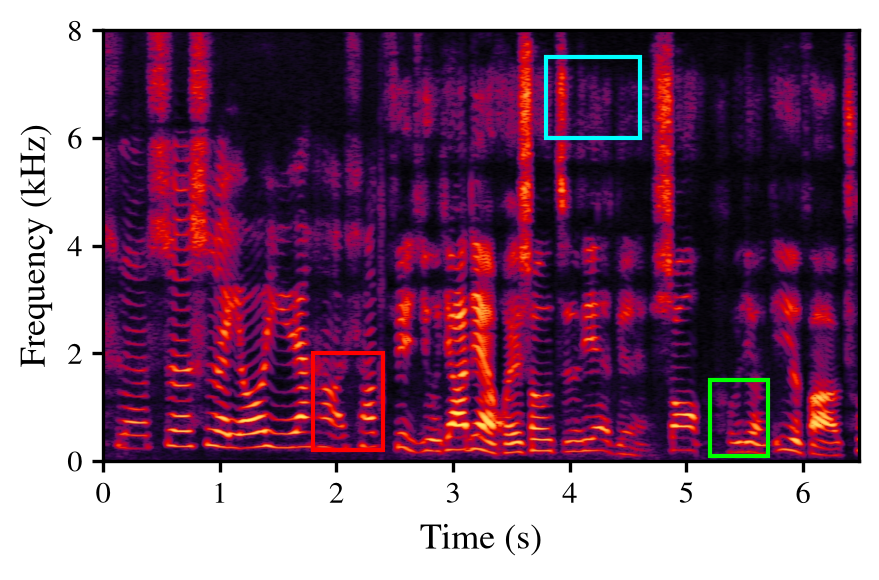}
  \centerline{\footnotesize (c) BBED}\medskip
\end{minipage}
\begin{minipage}{0.32\textwidth}
  \centering
  \includegraphics[width=\linewidth]{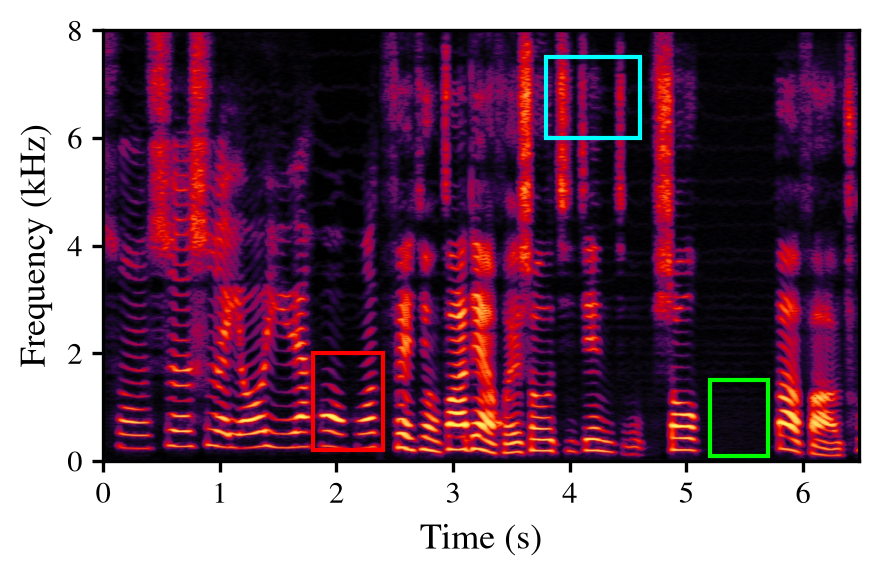}
  \centerline{\footnotesize (d) SB-VE}\medskip
\end{minipage}
\hfill
\begin{minipage}{0.32\textwidth}
  \centering
  \includegraphics[width=\linewidth]{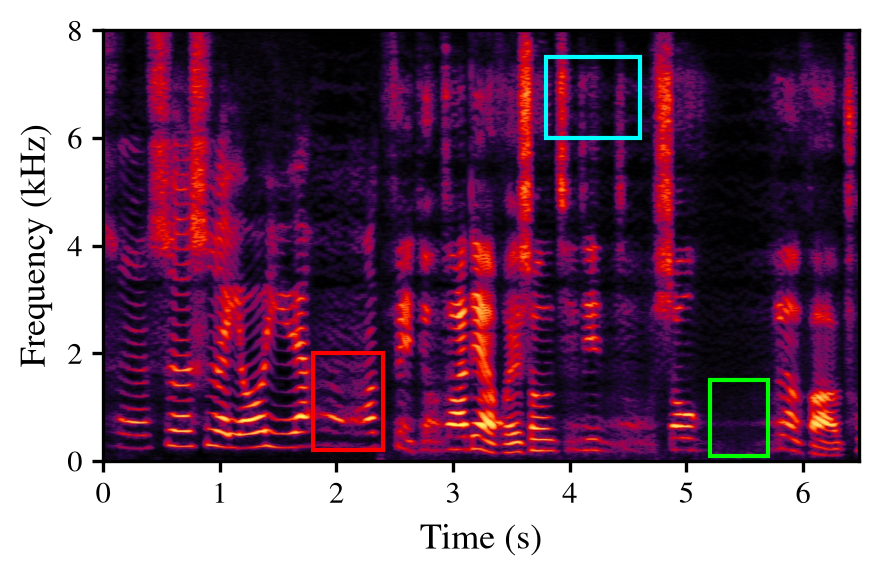}
  \centerline{\footnotesize (e) BB-RF}\medskip
\end{minipage}
\hfill
\begin{minipage}{0.32\textwidth}
  \centering
  \includegraphics[width=\linewidth]{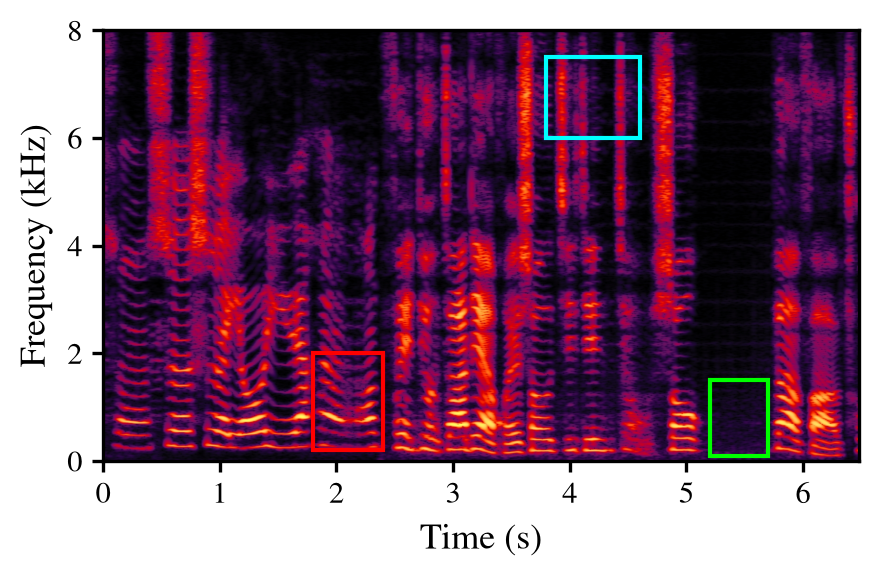}
  \centerline{\footnotesize (f) SB-RF (proposed)}\medskip
\end{minipage}
\caption{Spectrograms of enhanced speech from different generative SE methods on a Track B low-SNR example. SB-RF is visually closest to the clean reference and contains fewer noticeable artifacts in this example.}
\label{fig:spectrograms}
\end{figure*}

\noindent\textbf{Effect of inference steps.}
We further evaluate the performance of SB-RF across NFE settings (1, 5, 10) in Table~\ref{tab:simu_comparison2}. Although increasing the NFE to 5 or 10 leads to a slight improvement in signal fidelity metrics (SI-SDR rises from 10.4 to 10.7 dB) and marginal gains in DNSMOS, ESTOI remains stable at 0.70 and PESQ shows a degradation of up to 0.12. 
This phenomenon is consistent with the training objective, since the auxiliary losses in \eqref{eq:sbrf_loss} directly supervise the single-step reconstruction $\hat{\mathbf{x}} = \mathbf{x}_t - (t-\epsilon)\mathbf{v}_\theta$. With multiple Euler steps, the intermediate outputs at each sub-step are not directly constrained by these perceptual losses, so signal-level fidelity (SI-SDR) may improve while perceptual quality (PESQ) slightly degrades.
  
Given that increasing NFE introduces approximately a linear increase in computational costs without yielding substantial performance gains, NFE = 1 offers the optimal trade-off between efficiency and quality. 
Therefore, we use NFE = 1 as the main inference setting in this work. This result also supports the motivation for retaining the one-step RF inference rule in SB-RF.

\noindent\textbf{Effect of SB-based sampling.}
We also compare SB-RF with BB-RF to isolate the effect of using the SB time marginal instead of the perturbation kernel of the BB process, while keeping other components unchanged, including the backbone, losses, and inference rule. SB-RF improves over BB-RF across the metrics (e.g., +0.11 PESQ in Track A, +0.13 PESQ in Track B). Since both methods expose the model to stochastic intermediate states during training, this gain suggests that the improvement does not come from stochasticity alone, but also from the specific distribution of intermediate states.

\noindent\textbf{Spectrogram visualization.}
Fig.~\ref{fig:spectrograms} compares the spectrograms of clean, noisy, and enhanced audio produced by BBED, SB-VE, BB-RF, and SB-RF trained on Track B. The noisy input contains strong wind noise, which masks the harmonic structure of the target speech.
BBED removes substantial background noise but degrades harmonic details. In the green box, BBED introduces speech-like sounds absent from the clean reference. This is a common issue in generative models under low-SNR conditions, known as hallucination artifacts. 
SB-VE suppresses wind noise more thoroughly than BBED, but still attenuates high-frequency harmonics above 1.5\,kHz, as shown in the red box. 
Notably, both BBED and BB-RF fail to retain high-frequency components of voiced sounds in the cyan box, whereas SB-VE and SB-RF preserve them more clearly. This observation is consistent with our motivation for using the SB time marginal to improve the training distribution of the velocity field.
Overall, SB-RF is visually closest to the clean reference, preserving harmonic trajectories and high-frequency speech energy while suppressing noise effectively.
We note that this visualization presents a single representative example; the quantitative metrics in Tables~\ref{tab:vb_dmd_comparison} and~\ref{tab:simu_comparison2} support the observed trend.

\noindent\textbf{Ablation of auxiliary losses.}
Table~\ref{tab:ablation} examines the contribution of auxiliary losses. Removing both auxiliary losses (w/o $\mathcal{L}_{mel}$ and $\mathcal{L}_{pesq}$) decreases PESQ from 2.56 to 1.90. Nevertheless, compared with BBED and SB-VE in Table~\ref{tab:simu_comparison2}, this configuration achieves higher ESTOI and SI-SDR, while its PESQ (1.90) lies between the two baselines. This indicates that SB time-marginal sampling with the RF objective already provides an effective one-step solution.
Removing only $\mathcal{L}_{pesq}$ lowers PESQ to 2.14, while removing $\mathcal{L}_{mel}$ keeps PESQ nearly unchanged but reduces ESTOI, SI-SDR, and DNSMOS. These results show that the auxiliary losses help obtain a balanced one-step enhancement result. Such direct reconstruction supervision is particularly suitable for one-step RF, where the enhanced speech is obtained with a single reverse Euler step while avoiding back-propagation through a multi-step reverse process.
\begin{table}[t]
  \caption{Ablation study of training losses on Track B test set}
  \label{tab:ablation}
  \centering                                                           
  \begin{tabular}{|l|c|c|c|c|}
  \hline               
  \textbf{Loss configuration} & \textbf{PESQ} & \textbf{ESTOI} & \textbf{SI-SDR} & \textbf{DNSMOS} \\                  
  \hline
  SB-RF (full loss) & {2.56} & {0.70} & 10.4 & {3.41} \\
  w/o $\mathcal{L}_{mel}$ and $\mathcal{L}_{pesq}$ & 1.90 & 0.67 & {11.1} & 3.34 \\
  w/o $\mathcal{L}_{pesq}$ & 2.14 & 0.69 & 11.0 & 3.36 \\
  w/o $\mathcal{L}_{mel}$ & 2.57 & 0.68 & 9.8 & 3.37 \\
  \hline
  \end{tabular}
  \end{table}
\section{Conclusion}
In this paper, we present SB-RF, a framework that combines intermediate-state sampling from the analytic SB time marginal during training with the RF velocity-matching objective for one-step speech enhancement.
Experimental results demonstrate that, with one-step inference (NFE = 1), SB-RF achieves superior overall performance over the compared baselines on the standard VB-DMD benchmark and the simulated low-SNR test set. The comparison between BB-RF and SB-RF shows that, when the backbone, losses, and inference rule are fixed, replacing the BB perturbation kernel with the SB time marginal leads to a learned velocity field that yields better enhancement performance.
These results suggest that incorporating the SB time marginal into RF training improves the robustness of one-step generative speech enhancement while preserving inference efficiency.

Future work will focus on designing lightweight backbone architectures to reduce model size and computational cost, and exploring low-latency streaming implementations for real-time speech enhancement. We also plan to investigate the generalization of SB-RF to other audio tasks such as bandwidth extension and speech separation.
\clearpage
\bibliographystyle{IEEEtran}
\bibliography{myref}

@misc{chen2025meanflowse, 
    title        = {{MeanFlowSE}: one-step generative speech enhancement via {MeanFlow}},
    author       = {Zhu, Yike and Kang, Boyi and Wang, Ziqian and Li, Xingchen and Zhang, Zihan and Li, Wenjie and Xiao, Longshuai and Xue, Wei and Xie, Lei},
    year         = {2025},
    eprint       = {2509.23299},
    archivePrefix= {arXiv},
    primaryClass = {cs.SD},
    url          = {https://arxiv.org/abs/2509.23299},
    note         = {unpublished, arXiv:2509.23299}
}

@misc{chen2023schrodingerbridges, 
    title        = {{Schrodinger} bridges beat diffusion models on text-to-speech synthesis},
    author       = {Chen, Zehua and He, Guande and Zheng, Kaiwen and Tan, Xu and Zhu, Jun},
    year         = {2023},
    eprint       = {2312.03491},
    archivePrefix= {arXiv},
    primaryClass = {cs.LG},
    url          = {https://arxiv.org/abs/2312.03491},
    note         = {unpublished, arXiv:2312.03491}
}

@misc{voicerestore2025, 
    title        = {{VoiceRestore}: flow-matching transformers for speech recording quality restoration},
    author       = {Kirdey, Stanislav},
    year         = {2025},
    eprint       = {2501.00794},
    archivePrefix= {arXiv},
    primaryClass = {eess.AS},
    url          = {https://arxiv.org/abs/2501.00794},
    note         = {unpublished, arXiv:2501.00794}
}

@misc{compose2024, 
    title        = {Compose yourself: average-velocity flow matching for one-step speech enhancement},
    author       = {Yang, Gang and Lei, Yue and Tai, Wenxin and Wu, Jin and Chen, Jia and Zhong, Ting and Zhou, Fan},
    year         = {2025},
    eprint       = {2509.15952},
    archivePrefix= {arXiv},
    primaryClass = {cs.SD},
    url          = {https://arxiv.org/abs/2509.15952},
    note         = {unpublished, arXiv:2509.15952}
}

@inproceedings{de2022likelihood, 
    title        = {Likelihood training of {Schr{\"o}dinger} bridge using forward-backward {SDEs} theory},
    author       = {Chen, Tianrong and Liu, Guan-Horng and Theodorou, Evangelos A.},
    booktitle    = {Proc. International Conference on Learning Representations (ICLR)},
    year         = {2022}
}

@inproceedings{flowingstraighter2024,
    author       = {Cross, Mattias and Ragni, Anton},
    title        = {Flowing straighter with conditional flow matching for accurate speech enhancement},
    booktitle    = {Proc. 2nd ECAI Workshop on Machine Learning Meets Differential Equations: From Theory to Applications},
    series       = {Proceedings of Machine Learning Research},
    volume       = {277},
    pages        = {121--132},
    year         = {2025},
    publisher    = {PMLR}
}

@inproceedings{itu2001pesq, 
    author       = {Rix, A. W. and Beerends, J. G. and Hollier, M. P. and Hekstra, A. P.},
    title        = {Perceptual evaluation of speech quality ({PESQ}) -- a new method for speech quality assessment of telephone networks and codecs},
    booktitle    = {Proc. IEEE International Conference on Acoustics, Speech and Signal Processing (ICASSP)},
    pages        = {749--752},
    year         = {2001}
}

@inproceedings{lee2025flowse,
    author       = {Lee, Seonggyu and Cheong, Sein and Han, Sangwook and Shin, Jong Won},
    title        = {{FlowSE}: flow matching-based speech enhancement},
    booktitle    = {IEEE International Conference on Acoustics, Speech and Signal Processing (ICASSP)},
    publisher    = {IEEE},
    pages        = {1--5},
    year         = {2025}
}

@inproceedings{gerkmann2010empirical, 
    author       = {Gerkmann, Timo and Martin, Rainer},
    title        = {Empirical distributions of {DFT}-domain speech coefficients based on estimated speech variances},
    booktitle    = {Proc. International Workshop on Acoustic Echo and Noise Control (IWAENC)},
    pages        = {1--4},
    year         = {2010}
}

@inproceedings{braun2021consolidated, 
    author       = {Braun, Sebastian and Tashev, Ivan},
    title        = {A consolidated view of loss functions for supervised deep learning-based speech enhancement},
    booktitle    = {Proc. International Conference on Telecommunications and Signal Processing (TSP)},
    year         = {2021},
    pages        = {72--76}
}

@article{williamson2017time, 
    author       = {Williamson, Donald S. and Wang, DeLiang},
    title        = {Time-frequency masking in the complex domain for speech dereverberation and denoising},
    journal      = {IEEE/ACM Transactions on Audio, Speech, and Language Processing},
    volume       = {25},
    number       = {7},
    pages        = {1492--1501},
    year         = {2017}
}

@inproceedings{pascual2017segan, 
    author       = {Pascual, Santiago and Bonafonte, Antonio and Serr{\`a}, Joan},
    title        = {{SEGAN}: speech enhancement generative adversarial network},
    booktitle    = {Proc. {INTERSPEECH} 2017 -- 18\textsuperscript{th} Annual Conference of the International Speech Communication Association},
    year         = {2017},
    address      = {Stockholm, Sweden},
    month        = {{Aug.}},
    pages        = {3642--3646}
}

@article{luo2018conv,  
    author       = {Luo, Yi and Mesgarani, Nima},
    title        = {{Conv-TasNet}: surpassing ideal time-frequency magnitude masking for speech separation},
    journal      = {IEEE/ACM Transactions on Audio, Speech, and Language Processing},
    volume       = {27},
    number       = {8},
    year         = {2019},
    pages        = {1256--1266}
}

@inproceedings{leglaive2018variance,  
    author       = {Leglaive, Simon and Girin, Laurent and Horaud, Radu},
    title        = {A variance modeling framework based on variational autoencoders for speech enhancement},
    booktitle    = {Proc. International Workshop on Machine Learning for Signal Processing (MLSP)},
    year         = {2018},
    pages        = {1--6},
    organization = {IEEE}
}

@inproceedings{subramanian2019sergan, 
    author       = {Baby, Deepak and Verhulst, Sarah},
    title        = {{SERGAN}: speech enhancement using relativistic generative adversarial networks with gradient penalty},
    booktitle    = {IEEE International Conference on Acoustics, Speech and Signal Processing (ICASSP)},
    publisher    = {IEEE},
    year         = {2019},
    pages        = {106--110}
}

@inproceedings{fang2021variational, 
    author       = {Fang, Huajian and Carbajal, Guillaume and Wermter, Stefan and Gerkmann, Timo},
    title        = {Variational autoencoder for speech enhancement with a noise-aware encoder},
    booktitle    = {IEEE International Conference on Acoustics, Speech and Signal Processing (ICASSP)},
    year         = {2021},
    publisher    = {IEEE},
    pages        = {676--680},
    doi          = {10.1109/ICASSP39728.2021.9414060}
}

@inproceedings{lu2023mp,
    author       = {Lu, Ye-Xin and Ai, Yang and Ling, Zhen-Hua},
    title        = {{MP-SENet}: a speech enhancement model with parallel denoising of magnitude and phase spectra},
    booktitle    = {Proc. {INTERSPEECH} 2023 -- 24\textsuperscript{th} Annual Conference of the International Speech Communication Association},
    year         = {2023},
    month        = {{Aug.}},
    pages        = {3834--3838},
    address      = {Dublin, Ireland},
    doi          = {10.21437/Interspeech.2023-1441}
}

@inproceedings{li2024compression,
    author       = {Kumar, Rithesh and Seetharaman, Prem and Luebs, Alejandro and Kumar, Ishaan and Kumar, Kundan},
    title        = {High-fidelity audio compression with improved {RVQGAN}},
    booktitle    = {Advances in Neural Information Processing Systems (NeurIPS)},
    volume       = {36},
    pages        = {27980--27993},
    year         = {2023}
}

@inproceedings{chen2018schrodinger, 
    author       = {Bunne, Charlotte and Hsieh, Ya-Ping and Cuturi, Marco and Krause, Andreas},
    title        = {The {Schr{\"o}dinger} bridge between {Gaussian} measures has a closed form},
    booktitle    = {Proc. International Conference on Artificial Intelligence and Statistics (AISTATS)},
    pages        = {5802--5833},
    year         = {2023},
    publisher    = {PMLR}
}

@inproceedings{de2021diffusion, 
    author       = {De Bortoli, Valentin and Thornton, James and Heng, Jeremy and Doucet, Arnaud},
    title        = {Diffusion {Schr{\"o}dinger} bridge with applications to score-based generative modeling},
    booktitle    = {Advances in Neural Information Processing Systems (NeurIPS)},
    volume       = {34},
    pages        = {17695--17709},
    year         = {2021}
}

@inproceedings{lu2022conditional, 
    author       = {Lu, Yen-Ju and Wang, Zhong-Qiu and Watanabe, Shinji and Richard, Alexander and Yu, Cheng and Tsao, Yu},
    title        = {Conditional diffusion probabilistic model for speech enhancement},
    booktitle    = {IEEE International Conference on Acoustics, Speech and Signal Processing (ICASSP)},
    year         = {2022},
    publisher    = {IEEE},
    pages        = {7402--7406}
}

@article{richter2023speech, 
    author       = {Richter, Julius and Welker, Simon and Lemercier, Jean-Marie and Lay, Bunlong and Gerkmann, Timo},
    title        = {Speech enhancement and dereverberation with diffusion-based generative models},
    journal      = {IEEE/ACM Transactions on Audio, Speech, and Language Processing},
    volume       = {31},
    pages        = {2351--2364},
    year         = {2023}
}

@inproceedings{schneider2023reducing, 
    author       = {Lay, Bunlong and Welker, Simon and Richter, Julius and Gerkmann, Timo},
    title        = {Reducing the prior mismatch of stochastic differential equations for diffusion-based speech enhancement},
    booktitle    = {Proc. {INTERSPEECH} 2023 -- 24\textsuperscript{th} Annual Conference of the International Speech Communication Association},
    year         = {2023},
    month        = {{Aug.}},
    pages        = {3809--3813},
    doi          = {10.21437/Interspeech.2023-1445},
    address      = {Dublin, Ireland}
}

@inproceedings{larf2024,  
    author       = {Li, Zheng-Xiao and Inoue, Nakamasa},
    title        = {Locally aligned rectified flow model for speech enhancement toward single-step diffusion},
    booktitle    = {Proc. {INTERSPEECH} 2024 -- 25\textsuperscript{th} Annual Conference of the International Speech Communication Association},
    year         = {2024},
    month        = {{Sep.}},
    address      = {Kos, Greece},
    pages        = {2195--2199},
    doi          = {10.21437/Interspeech.2024-162}
}

@inproceedings{jukic2024schrodinger,  
    author       = {Juki{\'c}, Ante and Korostik, Roman and Balam, Jagadeesh and Ginsburg, Boris},
    title        = {{Schr{\"o}dinger} bridge for generative speech enhancement},
    booktitle    = {Proc. {INTERSPEECH} 2024 -- 25\textsuperscript{th} Annual Conference of the International Speech Communication Association},
    year         = {2024},
    month        = {{Sep.}},
    address      = {Kos, Greece},
    pages        = {1175--1179},
    doi          = {10.21437/Interspeech.2024-579}
}

@article{bronkhorst2002intelligibility, 
    author       = {Jensen, Jesper and Taal, Cees H.},
    title        = {An algorithm for predicting the intelligibility of speech masked by modulated noise maskers},
    journal    = {IEEE/ACM Transactions on Audio, Speech, and Language Processing},
    volume       = {24},
    number       = {11},
    year         = {2016},
    pages        = {2009--2022}
}

@inproceedings{leroux2019sdr,  
    author       = {Le Roux, Jonathan and Wisdom, Scott and Erdogan, Hakan and Hershey, John R.},
    title        = {{SDR} -- half-baked or well done?},
    booktitle    = {IEEE International Conference on Acoustics, Speech and Signal Processing (ICASSP)},
    year         = {2019},
    publisher    = {IEEE},
    month        = {{May}},
    pages        = {626--630}
}

@inproceedings{paliwal2020dnsmos, 
    author       = {Reddy, Chandan K. A. and Gopal, Vishak and Cutler, Ross},
    title        = {{DNSMOS}: a non-intrusive perceptual objective speech quality metric to evaluate noise suppressors},
    booktitle    = {IEEE International Conference on Acoustics, Speech and Signal Processing (ICASSP)},
    year         = {2021},
    publisher    = {IEEE},
    pages        = {6493--6497}
}

@misc{yamagishi2019vctk-corpus,
    title        = {{CSTR} {VCTK} corpus: {English} multi-speaker corpus for {CSTR} voice cloning toolkit (version 0.92)},
    author       = {Yamagishi, Junichi and Veaux, Christophe and MacDonald, Kirsten},
    year         = {2019},
    howpublished = {University of Edinburgh. The Centre for Speech Technology Research (CSTR)},
    doi          = {10.7488/ds/2645}
}

@inproceedings{thiemann2013demand, 
    author       = {Thiemann, Joachim and Ito, Nobutaka and Vincent, Emmanuel},
    title        = {The diverse environments multi-channel acoustic noise database ({DEMAND}): a database of multichannel environmental noise recordings},
    booktitle    = {Proceedings of Meetings on Acoustics},
    volume       = {19},
    number       = {1},
    pages        = {035081},
    year         = {2013},
    doi          = {10.1121/1.4799597}
}

@inproceedings{dubey2022icassp, 
    title        = {{ICASSP} 2022 deep noise suppression challenge},
    author       = {Dubey, Harishchandra and Gopal, Vishak and Cutler, Ross and Aazami, Ashkan and Matusevych, Sergiy and Braun, Sebastian and Eskimez, Sefik Emre and Thakker, Manthan and Yoshioka, Takuya and Gamper, Hannes and Aichner, Robert},
    booktitle    = {IEEE International Conference on Acoustics, Speech and Signal Processing (ICASSP)},
    year         = {2022},
    publisher    = {IEEE},
    month        = {{May}},
    pages        = {9271--9275},
    doi          = {10.1109/ICASSP43922.2022.9747230}
}

@inproceedings{ma2024wenetspeech4tts, 
    title        = {{WenetSpeech4TTS}: a 12,800-hour {Mandarin} {TTS} corpus for large speech generation model benchmark},
    author       = {Ma, Linhan and Guo, Dake and Song, Kun and Jiang, Yuepeng and Wang, Shuai and Xue, Liumeng and Xu, Weiming and Zhao, Huan and Zhang, Binbin and Xie, Lei},
    booktitle    = {Proc. {INTERSPEECH} 2024 -- 25\textsuperscript{th} Annual Conference of the International Speech Communication Association},
    year         = {2024},
    pages        = {1840--1844}
}

@misc{snyder2015musan, 
    title        = {{MUSAN}: a music, speech, and noise corpus},
    author       = {Snyder, David and Chen, Guoguo and Povey, Daniel},
    year         = {2015},
    eprint       = {1510.08484},
    archivePrefix= {arXiv},
    primaryClass = {cs.SD},
    url          = {https://arxiv.org/abs/1510.08484},
    note         = {unpublished, arXiv:1510.08484}
}

@inproceedings{bu2017aishell1, 
    title        = {{AISHELL-1}: an open-source {Mandarin} speech corpus and a speech recognition baseline},
    author       = {Bu, Hui and Du, Jiayu and Na, Xingyu and Wu, Bengu and Zheng, Hao},
    booktitle    = {2017 Conference of the Oriental Chapter of the International Committee for the Coordination and Standardization of Speech Databases and Assessment Techniques (O-COCOSDA)},
    year         = {2017},
    pages        = {1--5},
    doi          = {10.1109/ICSDA.2017.8384449}
}

@inproceedings{panayotov2015librispeech, 
    title        = {{LibriSpeech}: an {ASR} corpus based on public domain audio books},
    author       = {Panayotov, Vassil and Chen, Guoguo and Povey, Daniel and Khudanpur, Sanjeev},
    booktitle    = {IEEE International Conference on Acoustics, Speech and Signal Processing (ICASSP)},
    year         = {2015},
    publisher    = {IEEE},
    pages        = {5206--5210},
    doi          = {10.1109/ICASSP.2015.7178964}
}

@inproceedings{wichern2019wham, 
    title        = {{WHAM!}: extending speech separation to noisy environments},
    author       = {Wichern, Gordon and Antognini, Joe and Flynn, Michael and Zhu, Licheng Richard and McQuinn, Emmett and Crow, Dwight and Manilow, Ethan and Le Roux, Jonathan},
    booktitle    = {Proc. {INTERSPEECH} 2019 -- 20\textsuperscript{th} Annual Conference of the International Speech Communication Association},
    year         = {2019},
    pages        = {1368--1372}
}

@article{leonard2012from,
    title        = {From the {Schr{\"o}dinger} problem to the {Monge-Kantorovich} problem},
    author       = {L{\'e}onard, Christian},
    year         = {2012},
    journal      = {Journal of Functional Analysis},
    volume       = {262},
    number       = {4},
    pages        = {1879--1920},
    doi          = {10.1016/j.jfa.2011.11.026}
}

@inproceedings{song2021scorebased,
    title        = {Score-based generative modeling through stochastic differential equations},
    author       = {Song, Yang and Sohl-Dickstein, Jascha and Kingma, Diederik P. and Kumar, Abhishek and Ermon, Stefano and Poole, Ben},
    booktitle    = {Proc. International Conference on Learning Representations (ICLR)},
    year         = {2021}
}

@article{vincent2011smdae,
    title        = {A connection between score matching and denoising autoencoders},
    author       = {Vincent, Pascal},
    journal      = {Neural Computation},
    volume       = {23},
    number       = {7},
    pages        = {1661--1674},
    year         = {2011}
}

@article{bishop1995tikhonov,
    title        = {Training with noise is equivalent to {Tikhonov} regularization},
    author       = {Bishop, Chris M.},
    journal      = {Neural Computation},
    volume       = {7},
    number       = {1},
    pages        = {108--116},
    year         = {1995}
}

@inproceedings{liu2023flow,
    title        = {Flow straight and fast: learning to generate and transfer data with rectified flows},
    author       = {Liu, Xingchao and Gong, Chengyue and Liu, Qiang},
    booktitle    = {Proc. International Conference on Learning Representations (ICLR)},
    year         = {2023}
}

\end{document}